\documentclass[sigconf]{acmart}
\AtBeginDocument{%
  \providecommand\BibTeX{{%
    \normalfont B\kern-0.5em{\scshape i\kern-0.25em b}\kern-0.8em\TeX}}}

\usepackage{listings}

\usepackage{graphics}

\usepackage{pgfplots}
\usepackage{subcaption}
\usepackage{makecell}
\newtheorem{definition}{Definition}[section]

\usepackage{color} 

\def\tsc#1{\csdef{#1}{\textsc{\lowercase{#1}}\xspace}}
\tsc{WGM}
\tsc{QE}

\acmConference[BIR '23]{ECIR 2023}{April 02,
  2023}{Dublin, Ireland}
\begin{document}

\title{RecBaselines2023: a new dataset for choosing baselines for recommender models}

\author{Veronika Ivanova}
\email{veronika.ivanova88@yandex.ru}
\orcid{0009-0009-6864-392X}

\affiliation{%
  \institution{National Research University Higher School of Economics}
  \streetaddress{Myasnitskaya Ulitsa, 20}
  \city{Moscow}
  \country{Russian Federation}
  \postcode{101000}
}

\author{Oleg Lashinin}
\email{o.a.lashinin@tinkoff.ru}
\orcid{0000-0001-8894-9592}
\affiliation{%
  \institution{Tinkoff}
  \streetaddress{2-Ya Khutorskaya Ulitsa, 38A, bld. 26}
  \city{Moscow}
  \country{Russian Federation}
  \postcode{117198}
}

\author{Marina Ananyeva}
\email{m.ananyeva@tinkoff.ru}
\orcid{0000-0002-9885-2230}
\affiliation{%
  \institution{National Research University Higher School of Economics}
  \streetaddress{Myasnitskaya Ulitsa, 20}
  \city{Moscow}
  \country{Russian Federation}
  \postcode{101000}
}

\author{Sergey Kolesnikov}
\email{scitator@gmail.com}
\orcid{0000-0002-4820-987X}
\affiliation{%
  \institution{Tinkoff}
  \streetaddress{2-Ya Khutorskaya Ulitsa, 38A, bld. 26}
  \city{Moscow}
  \country{Russian Federation}
  \postcode{117198}
}






\renewcommand{\shortauthors}{Ivanova, Lashinin, Ananyeva and Kolesnikov}

\begin{abstract}
The number of proposed recommender algorithms continues to grow. The authors propose new approaches and compare them with existing models, called baselines. Due to the large number of recommender models, it is difficult to estimate which algorithms to choose in the article. To solve this problem, we have collected and published a dataset containing information about the recommender models used in 903 papers, both as baselines and as proposed approaches. This dataset can be seen as a typical dataset with interactions between papers and previously proposed models. In addition, we provide a descriptive analysis of the dataset and highlight possible challenges to be investigated with the data. Furthermore, we have conducted extensive experiments using a well-established methodology to build a good recommender algorithm under the dataset. Our experiments show that the selection of the best baselines for proposing new recommender approaches can be considered and successfully solved by existing state-of-the-art collaborative filtering models. Finally, we discuss limitations and future work.

\end{abstract}





\maketitle

\section{Introduction}

There is an increasing number of publications in the field of recommender systems. Authors need to evaluate the performance of the proposed model against reference models to demonstrate its efficiency. Reference models are usually referred to as baselines. However, there are no rigid guidelines that define a comprehensive list of essential baselines. Inaccurate selection of baselines can lead to incorrect conclusions about the performance of the proposed model. Subsequent papers \cite{ferrari2021troubling} on the reproducibility and progress of existing work have demonstrated this fact. For example, in two recent papers \cite{lin2019neural, yang2019critically}, the authors report that for a particular information retrieval task, some non-neural methods outperform recent neural methods. In 2016, Kharazmi et al. \cite{kharazmi2016examining} examined previous work on IR and found a tendency to select weak baselines for comparative experiments. In the field of recommender systems, the empirical analyses of session-based recommendation papers showed that sometimes almost trivial methods can outperform the latest neural methods \cite{ludewig2018evaluation, ludewig2019performance}. One of the reasons for such disappointing performance of novel models is the poor choice of baselines, which gives the illusion of better results. Another consequence of not choosing appropriate baselines for a new algorithm is that the proposed paper may be rejected \cite{bedi2022did}. Thus, the choice of baselines is currently one of the major issues in recommender systems research \cite{ferrari2019we, beel2016paper}. 

With accurate baseline selection, the development of recommender systems can progress more quickly. Both researchers and practitioners are faced with an increasing number of models to select for their experiments in order to consider relevant baselines. However, the number of baselines included in the paper is limited for the following reasons. First, more baselines require more computational time. The recent success of deep learning forces the inclusion of complicated algorithms as baselines. Therefore, some researchers cannot afford to choose effective hyperparameters sufficiently \cite{ferrari2019we}. Second, some papers with new recommender algorithms do not have the source code of the implementation \cite{ferrari2019we}. This may lead to poor performance of third party implementations \cite{petrov2022systematic, ferrari2019we}. Finally, due to space limitations, a paper cannot include too many baselines. Thus, it is common practice to study the performance of only 3-7 baselines against the newly proposed method. The problem of selecting a few relevant items from a large set is a well-known task and can be solved by recommender systems \cite{overload}. To the best of our knowledge, there is no open source dataset that can be used as a basis for developing the recommender baseline suggestion system.

It is important to note that the baseline recommendations can be applied to other areas of machine learning, such as natural language processing, computer vision, time series prediction, and others. However, in this paper we focus only on recommender systems.

In this paper we describe a process for collecting a novel dataset called \text{RecBaselines2023}. It can be considered as a classical dataset with interactions between papers and baselines. In addition, we present the results of experiments that have been performed on RecBaselines2023. Our results show potential advantages of our experiments and open new research directions. Specifically, the main contribution of this paper can be listed as follows:

\begin{itemize}
    \item We have created a new open source dataset called RecBaseline2023\footnote{We are releasing an online version of the dataset: \url{https://github.com/fotol1/recbaselines2023}. }
    for selecting baselines for experiments on recommender models.  We examined 1009 papers for the collection process. After preprocessing, RecBaseline2023 contains information on 363 baselines used in 903 articles published between 2010 and 2022. We also provide a data collection procedure and descriptive statistics.
    \item We discuss that the problem of baseline selection can be solved by collaborative filtering approaches. We then compare the baseline ranking quality of seven state-of-the-art top-N recommender models on RecBaseline2023. The results show that this problem can be effectively solved by selected algorithms. 
    \item We describe a scenario where a partial list of baselines needs to be completed. The list is given to collaborative filtering approaches that recommend baselines based on the list of methods already used. Some other possible use cases of RecBaseline2023 are also mentioned.
\end{itemize}

\section{Related work}

The problem of choosing baselines for research experiments in machine learning is not well studied. A similar problem is citation recommendations. This direction aims at suggesting other papers to cite.

The two main classes of citation recommendations are content-based and collaborative filtering \cite{beel2016paper}. The content-based methods use textual elements such as abstract and title or metadata elements such as authors. In \cite{bhagavatula2018content}, the authors proposed a content-based approach requiring only textual features and collected the OpenCorpus dataset of 7 million articles. The literature graph was created in \cite{ammar2018construction} using nodes for articles, authors and scientific concepts. Collaborative filtering methods are based on comparing similarities between articles. Liu et al. \cite{liu2015context} measured the cosine similarity of article vectors and created article vectors based on co-occurrence in the same citation list. The same concept was used by Haruna et al. \cite{haruna2017collaborative}. However, they considered the references and citations of the target paper and mined the hidden associations between them using paper-citation relationships. Later, by improving the similarity calculation, this approach was further developed in \cite{sakib2020collaborative}.
 
Although we know what to cite, it is not clear whether the recommended paper should be used as a baseline. Therefore, researchers are also working on more specialised tasks such as tag or baseline recommendations. The task of tag recommendation has been successfully studied by Wang et al. \cite{wang2013collaborative}.  They used the collaborative topic regression model. The authors sampled items from the CiteULike dataset, including abstracts, titles and tags for each article. Bedi et al. \cite{bedi2022did} introduced the task of identifying the papers used as baselines in a given scientific article. The author formulated it as a reference classification problem on a developed ACL anthology corpus dataset, where about 2000 papers were selected and manually annotated.

However, research article datasets are not specifically designed for the task of selecting baselines for recommender system experiments. We hope that our dataset will help to fill this gap and provide researchers with a practical approach to selecting baseline models for their research.

\section{Dataset}

\textbf{Collection}. We added several common recommendation tasks to our dataset, including the traditional top-n, next-item and next-basket recommendations. These tasks were used as the basis for the data collection. For each task there are well-established and highly cited baselines, some of which are listed in \autoref{tab:linebaseline}. Note that there are no strict guidelines in recommender systems research as to which baselines should be used for each of the above tasks. Therefore, we cannot guarantee that other algorithms cannot be used to complement the list of commonly used algorithms. However, the approaches listed in \autoref{tab:linebaseline} have many citations, which is appropriate for the starting point of data collection.

\begin{table}[t!]
    \caption{The table shows the corresponding highly cited baselines in each of the three recommender tasks. We use these as a starting point for our data collection. Citation counts are valid as of 27 March 2023.}
    \label{tab:linebaseline}
    \centering
     \resizebox{\columnwidth}{!}{%
    \begin{tabular}{ll}
    \hline
        \textbf{Recommender task} & \textbf{Mandatory baselines (Citation counts) }\\
    \hline
        conventional top-N &  BPR-MF (5297) \cite{bpr}, WMF (3688) \cite{wmf}, MultVAE (844) \cite{multivae}, LightGCN (1245) \cite{lightgcn}\\  
        next-item& GRU4Rec (2152) \cite{gru4rec}, SASRec (1146) \cite{sasrec}, BERT4Rec (836) \cite{sun2019bert4rec}\\
        next-basket& TIFUKNN (56) \cite{tifuknn}, RepeatNet (189) \cite{repeatnet}\\

    \hline
    \end{tabular}
    }
\end{table}

\begin{table}[t]
\caption{An example of a row in the RecBaselines2023 dataset. This file is called \textbf{before\_preprocessing.csv} and is available online.}
\label{tab:datasetdesc}
\centering
 \resizebox{\columnwidth}{!}{
\begin{tabular}{lll}
\hline
    \textbf{Column} & \textbf{Description }&\textbf{Example}\\
\hline
    Paper\_id & Unique paper identification&12\\
    URL & WEB link to a paper&arxiv.org/pdf/1809.07053.pdf\\
    Title & Paper title& \makecell{NAIS: Neural Attentive Item \\ Similarity Model for Recommendation}\\
    Year & A year of publishing &2018\\
    Baselines & List of used recommender algorithms  &MF;MLP;FISM;NAIS\\
\hline
\end{tabular}
}
\end{table}

\begin{table}[t]

    \caption{Examples of different names for the same algorithms}
    \label{tab:misnamings}
    \centering
    \begin{tabular}{ll}
    \hline
        \textbf{Approach} & \textbf{Occurring names }\\
    \hline
        POP & MostPop, Popular, TopPopular\\
        BPR-MF \cite{bpr} & BPR, BPR-MF, MF-BPR\\
        NCF \cite{neumf} & NeuCF, NeuMF, NCF \\
        Mult-VAE \cite{multvae}& Mult-VAE, Multi-VAE, VAE-CF\\
        
    \hline
    \end{tabular}
\end{table}

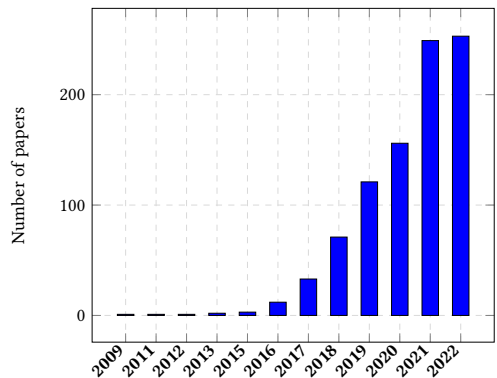
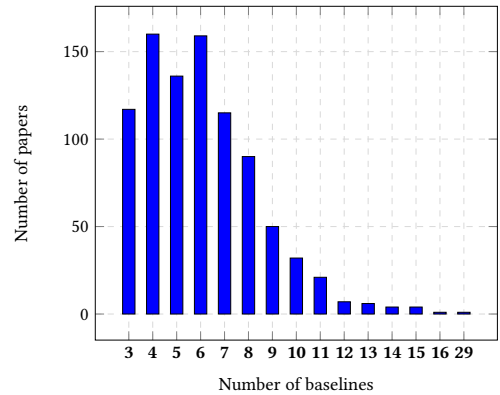
\begin{figure*}[t!]
\begin{subfigure}{.45\linewidth}
\centering
\begin{tikzpicture}[scale=0.78]
\begin{axis}[
    symbolic x coords={\textbf{2009},\textbf{2011},\textbf{2012},\textbf{2013},\textbf{2015},\textbf{2016},\textbf{2017},\textbf{2018},\textbf{2019},\textbf{2020},\textbf{2021},\textbf{2022}},
      xtick=data,
      grid=major,
      grid style={dashed,gray!30},
      bar width=8,
      x tick label style={rotate=45,anchor=east},
      ylabel=Number of papers ,
      ]
    \addplot[ybar,fill=blue] coordinates {	 (\textbf{2009}, 1)	 (\textbf{2011},1) (\textbf{2012}, 1) (\textbf{2013}, 2)	 	 (\textbf{2015}, 3)	 (\textbf{2016}, 12)	 (\textbf{2017}, 33)	 (\textbf{2018}, 71)	 (\textbf{2019}, 121)    (\textbf{2020}, 156) (\textbf{2021}, 249)    (\textbf{2022}, 253)};
\end{axis}
\end{tikzpicture}
\caption{Distribution of the number of papers by
the year of publication.}
\label{fig:years}
\end{subfigure}
\hfill
\begin{subfigure}{.45\linewidth}
\centering
\begin{tikzpicture}[scale=0.78]
\begin{axis}[
    symbolic x coords={
\textbf{3},\textbf{4},\textbf{5},\textbf{6},\textbf{7},\textbf{8},\textbf{9},\textbf{10},\textbf{11},\textbf{12},\textbf{13},\textbf{14},\textbf{15},\textbf{16},\textbf{29}},
      xtick=data,
      grid=major,
      grid style={dashed,gray!30},
      bar width=6,
      xlabel=Number of baselines,
      ylabel=Number of papers,
      ]
    \addplot[ybar,fill=blue] coordinates {
	 (\textbf{3}, 117)	 (\textbf{4}, 160)	 (\textbf{5}, 136)	 (\textbf{6}, 159)	 (\textbf{7}, 115)	 (\textbf{8}, 90)	 (\textbf{9}, 50)	 (\textbf{10}, 32)	 (\textbf{11}, 21)	 (\textbf{12}, 7)	 (\textbf{13}, 6)	 (\textbf{14}, 4)	 (\textbf{15}, 4)	 (\textbf{16}, 1)	 (\textbf{29}, 1) };
\end{axis}
\end{tikzpicture}
\caption{Distribution of the number of papers over the number of algorithms per one paper.}
\label{fig:baseline_number}
\end{subfigure}
\caption{Dataset statistics}
\label{fig:dataset_dist12}
\end{figure*}

To collect our dataset, we took the following steps:
\begin{enumerate} 
  \item For each model from \autoref{tab:linebaseline}, we obtain the list of papers that cited the model in Google Scholar \cite{jacso2005google}. If a paper included experiments with the model, we included it. We did not include papers with experiments on related problems (such as link prediction or matrix completion, explanation generation). In addition, papers without experiments are not included in the dataset. Note that a paper could cite more than one baseline model of \autoref{tab:linebaseline}. Duplicate papers were later filtered out of the dataset during pre-processing. Once we had gone through all the citations of models in \autoref{tab:linebaseline}, we continued to process citations of papers that had already been added. This was all done manually by the authors of the paper over the period of one month. 
  \item Information about each paper collected to build our dataset is presented in \autoref{tab:datasetdesc}. Each row contains a paper id, URL, paper title, year of publication and a list of recommender models used. The URL and year of publication are taken from the Google Scholar page, while the paper title and list of baselines are taken from the paper itself.
\end{enumerate}

After removing duplicates, we obtained the dataset with 1009 papers and 2187 baselines.  A large number of baseline models were only included in one or two papers. 

\textbf{Preprocessing}. A number of steps were taken to preprocess the data for future research:
\begin{enumerate}
    \item In some papers, popular models are presented under different names. This is most likely due to space limitations or different names for algorithms that were even proposed in the original paper. For example, the authors of the article Neural Collaborative Filtering (NCF) \cite{neumf} used a different name, NeuMF, in their experiments. As a result, the cited articles include both NCF and NeuMF. We try to condense common cases and list them in \autoref{tab:misnamings}.  To resolve this inconsistency, we have replaced the multiple names of a model with a single option.
    \item Some papers modify methods slightly and report different variations of the same methods. For example, the authors introduce three new loss functions in \cite{gao2022mcl} and apply them to different methods such as NeuMF \cite{neumf}, CML \cite{cml} and LightGCN \cite{lightgcn}. Considering three losses for each of the three models makes our dataset more sparse. To avoid this problem, the preprocessed version of RecBaselnies2023 contains only the main algorithms without any specified modifications.
    \item To replace rare baselines and papers with extremely few baselines, we then iteratively filtered the dataset until there were only papers with three or more baselines and each baseline was present in three or more papers. The resulting statistics for the filtered dataset can be found in \autoref{tab:statistics}.

\end{enumerate}

\begin{table}[t]
  \caption{Descriptive statistics of RecBaselines2023 dataset.}
  \label{tab:statistics}
  \centering
  \begin{tabular}{lllll}
    \toprule
    Stage &\# papers & \# models & \# interactions & density\\
    \midrule
    Collection&1009 & 2188 & 7748& 0.3\% \\
    Preprocessing&903 & 363 & 5467 & 1.6\% \\
    \bottomrule
  \end{tabular}
\end{table}

\textbf{Statistics}. We briefly present some statistics from the collected dataset. The main characteristics such as number of papers, number of models, number of interactions and density are presented in \autoref{tab:statistics}.

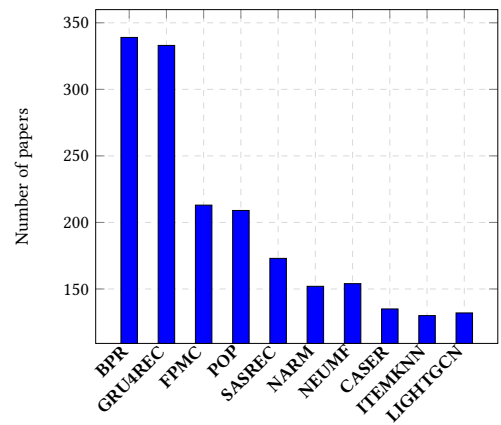
\begin{figure}
\centering
\begin{tikzpicture}[scale=0.78]
\begin{axis}[
    symbolic x coords={\textbf{BPR},\textbf{GRU4REC},\textbf{FPMC},\textbf{POP},\textbf{SASREC},\textbf{NARM},\textbf{NEUMF},\textbf{CASER},\textbf{ITEMKNN},\textbf{LIGHTGCN}},
      xtick=data,
      grid=major,
      grid style={dashed,gray!30},
      bar width=8,
      x tick label style={rotate=45,anchor=east},
      ylabel=Number of papers ,
      ]
    \addplot[ybar,fill=blue] coordinates {	 (\textbf{BPR}, 339)	 (\textbf{GRU4REC},333) (\textbf{FPMC}, 213) (\textbf{POP}, 209)	 	 (\textbf{SASREC}, 173)	 (\textbf{NEUMF}, 154)	 (\textbf{NARM}, 152)	 (\textbf{CASER}, 135)	 (\textbf{LIGHTGCN}, 132)(\textbf{ITEMKNN}, 130)    };
\end{axis}
\end{tikzpicture}
\caption{Distribution of the number of papers for the top-10 most popular baselines, included in our dataset.}
\label{fig:top_popular}
\label{fig:dataset_dist3}
\end{figure}

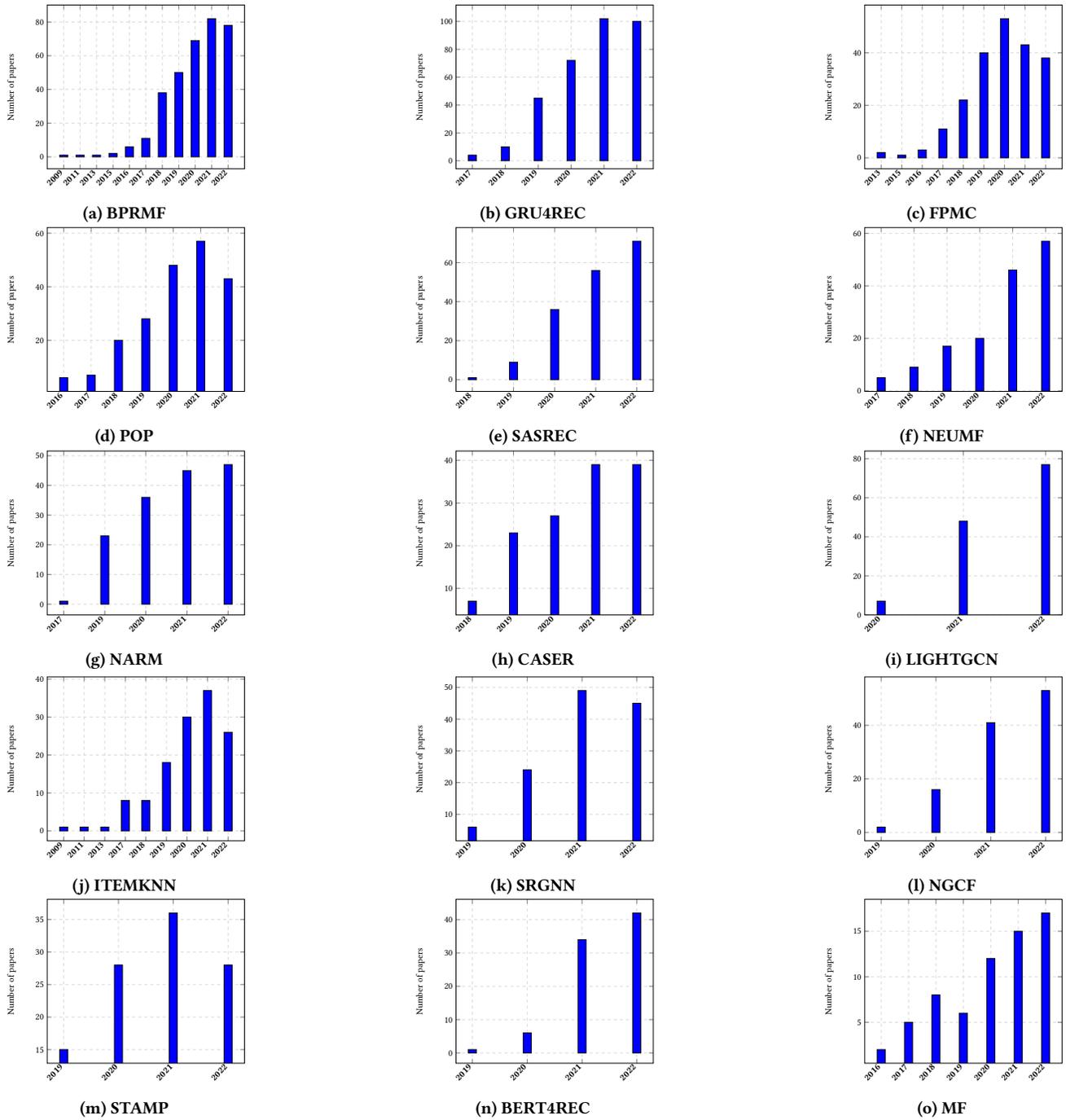
\begin{figure*}[t!]
\begin{subfigure}{.25\linewidth}
\centering
\begin{tikzpicture}[scale=0.47]
\begin{axis}[
    symbolic x coords={\textbf{2009},\textbf{2011},\textbf{2013},\textbf{2015},\textbf{2016},\textbf{2017},\textbf{2018},\textbf{2019},\textbf{2020},\textbf{2021},\textbf{2022}},
      xtick=data,
      grid=major,
      grid style={dashed,gray!30},
      bar width=8,
      x tick label style={rotate=45,anchor=east},
      ylabel=Number of papers ,
      ]
    \addplot[ybar,fill=blue] coordinates {	 (\textbf{2009}, 1)	 (\textbf{2011},1)  (\textbf{2013}, 1)	 	 (\textbf{2015}, 2)	 (\textbf{2016}, 6)	 (\textbf{2017}, 11)	 (\textbf{2018}, 38)	 (\textbf{2019}, 50)    (\textbf{2020}, 69) (\textbf{2021}, 82)    (\textbf{2022}, 78)};
\end{axis}
\end{tikzpicture}
\caption{BPRMF}
\label{fig:bprmf}
\end{subfigure}
\hfill
\begin{subfigure}{.25\linewidth}
\centering
\begin{tikzpicture}[scale=0.47]
\begin{axis}[
    symbolic x coords={\textbf{2017},\textbf{2018},\textbf{2019},\textbf{2020},\textbf{2021},\textbf{2022}},
      xtick=data,
      grid=major,
      grid style={dashed,gray!30},
      bar width=8,
      x tick label style={rotate=45,anchor=east},
      ylabel=Number of papers ,
      ylabel=Number of papers ,
      ]
    \addplot[ybar,fill=blue] coordinates {(\textbf{2017}, 4)	 (\textbf{2018}, 10)	 (\textbf{2019}, 45)    (\textbf{2020}, 72) (\textbf{2021}, 102)    (\textbf{2022}, 100)};
\end{axis}
\end{tikzpicture}
\caption{GRU4REC}
\label{fig:gru4rec}
\end{subfigure}
\hfill
\begin{subfigure}{.25\linewidth}
\centering
\begin{tikzpicture}[scale=0.47]
\begin{axis}[
    symbolic x coords={\textbf{2013},\textbf{2015},\textbf{2016},\textbf{2017},\textbf{2018},\textbf{2019},\textbf{2020},\textbf{2021},\textbf{2022}},
      xtick=data,
      grid=major,
      grid style={dashed,gray!30},
      bar width=8,
      x tick label style={rotate=45,anchor=east},
      ylabel=Number of papers ,
      ]
    \addplot[ybar,fill=blue] coordinates {(\textbf{2013}, 2)	 	 (\textbf{2015}, 1)	 (\textbf{2016}, 3)	 (\textbf{2017}, 11)	 (\textbf{2018}, 22)	 (\textbf{2019}, 40)    (\textbf{2020}, 53) (\textbf{2021}, 43)    (\textbf{2022}, 38)};
\end{axis}
\end{tikzpicture}
\caption{FPMC}
\label{fig:fpmc}
\end{subfigure}
\hfill
\begin{subfigure}{.25\linewidth}
\centering
\begin{tikzpicture}[scale=0.47]
\begin{axis}[
    symbolic x coords={\textbf{2016},\textbf{2017},\textbf{2018},\textbf{2019},\textbf{2020},\textbf{2021},\textbf{2022}},
      xtick=data,
      grid=major,
      grid style={dashed,gray!30},
      bar width=8,
      x tick label style={rotate=45,anchor=east},
      ylabel=Number of papers ,
      ylabel=Number of papers ,
      ]
    \addplot[ybar,fill=blue] coordinates {(\textbf{2016}, 6)(\textbf{2017}, 7)	 (\textbf{2018}, 20)	 (\textbf{2019}, 28)    (\textbf{2020}, 48) (\textbf{2021}, 57)    (\textbf{2022}, 43)};
\end{axis}
\end{tikzpicture}
\caption{POP}
\label{fig:pop}
\end{subfigure}
\hfill
\begin{subfigure}{.25\linewidth}
\centering
\begin{tikzpicture}[scale=0.47]
\begin{axis}[
    symbolic x coords={\textbf{2018},\textbf{2019},\textbf{2020},\textbf{2021},\textbf{2022}},
      xtick=data,
      grid=major,
      grid style={dashed,gray!30},
      bar width=8,
      x tick label style={rotate=45,anchor=east},
      ylabel=Number of papers ,
      ]
    \addplot[ybar,fill=blue] coordinates {(\textbf{2018}, 1)	 (\textbf{2019}, 9)    (\textbf{2020}, 36) (\textbf{2021}, 56)    (\textbf{2022}, 71)};
\end{axis}
\end{tikzpicture}
\caption{SASREC}
\label{fig:sasrec}
\end{subfigure}
\hfill
\begin{subfigure}{.25\linewidth}
\centering
\begin{tikzpicture}[scale=0.47]
\begin{axis}[
    symbolic x coords={\textbf{2017},\textbf{2018},\textbf{2019},\textbf{2020},\textbf{2021},\textbf{2022}},
      xtick=data,
      grid=major,
      grid style={dashed,gray!30},
      bar width=8,
      x tick label style={rotate=45,anchor=east},
      ylabel=Number of papers ,
      ]
    \addplot[ybar,fill=blue] coordinates {(\textbf{2017}, 5)	 (\textbf{2018}, 9)	 (\textbf{2019}, 17)    (\textbf{2020}, 20) (\textbf{2021}, 46)    (\textbf{2022}, 57)};
\end{axis}
\end{tikzpicture}
\caption{NEUMF}
\label{fig:neumf}
\end{subfigure}
\hfill
\begin{subfigure}{.25\linewidth}
\centering
\begin{tikzpicture}[scale=0.47]
\begin{axis}[
    symbolic x coords={\textbf{2017},\textbf{2019},\textbf{2020},\textbf{2021},\textbf{2022}},
      xtick=data,
      grid=major,
      grid style={dashed,gray!30},
      bar width=8,
      x tick label style={rotate=45,anchor=east},
      ylabel=Number of papers ,
      ylabel=Number of papers ,
      ]
    \addplot[ybar,fill=blue] coordinates {(\textbf{2017}, 1)	 (\textbf{2019}, 23)    (\textbf{2020}, 36) (\textbf{2021}, 45) (\textbf{2022}, 47)};
\end{axis}
\end{tikzpicture}
\caption{NARM}
\label{fig:narm}
\end{subfigure}
\hfill
\begin{subfigure}{.25\linewidth}
\centering
\begin{tikzpicture}[scale=0.47]
\begin{axis}[
    symbolic x coords={\textbf{2018},\textbf{2019},\textbf{2020},\textbf{2021},\textbf{2022}},
      xtick=data,
      grid=major,
      grid style={dashed,gray!30},
      bar width=8,
      x tick label style={rotate=45,anchor=east},
      ylabel=Number of papers ,
      ylabel=Number of papers ,
      ]
    \addplot[ybar,fill=blue] coordinates {(\textbf{2018}, 7)	 (\textbf{2019}, 23)    (\textbf{2020}, 27) (\textbf{2021}, 39)    (\textbf{2022}, 39)};
\end{axis}
\end{tikzpicture}
\caption{CASER}
\label{fig:caser}
\end{subfigure}
\hfill
\begin{subfigure}{.25\linewidth}
\centering
\begin{tikzpicture}[scale=0.47]
\begin{axis}[
    symbolic x coords={\textbf{2020},\textbf{2021},\textbf{2022}},
      xtick=data,
      grid=major,
      grid style={dashed,gray!30},
      bar width=8,
      x tick label style={rotate=45,anchor=east},
      ylabel=Number of papers ,
      ylabel=Number of papers ,
      ]
    \addplot[ybar,fill=blue] coordinates { (\textbf{2020}, 7) (\textbf{2021}, 48)    (\textbf{2022}, 77)};
\end{axis}
\end{tikzpicture}
\caption{LIGHTGCN}
\label{fig:lgcn}
\end{subfigure}
\hfill
\begin{subfigure}{.25\linewidth}
\centering
\begin{tikzpicture}[scale=0.47]
\begin{axis}[
    symbolic x coords={\textbf{2009},\textbf{2011},\textbf{2013},\textbf{2017},\textbf{2018},\textbf{2019},\textbf{2020},\textbf{2021},\textbf{2022}},
      xtick=data,
      grid=major,
      grid style={dashed,gray!30},
      bar width=8,
      x tick label style={rotate=45,anchor=east},
      ylabel=Number of papers ,
      ]
    \addplot[ybar,fill=blue] coordinates {	 (\textbf{2009}, 1)	 (\textbf{2011},1)  (\textbf{2013}, 1) (\textbf{2017}, 8)	 (\textbf{2018}, 8)	 (\textbf{2019}, 18)    (\textbf{2020}, 30) (\textbf{2021}, 37)    (\textbf{2022}, 26)};
\end{axis}
\end{tikzpicture}
\caption{ITEMKNN}
\label{fig:itemknn}
\end{subfigure}
\hfill
\begin{subfigure}{.25\linewidth}
\centering
\begin{tikzpicture}[scale=0.47]
\begin{axis}[
    symbolic x coords={\textbf{2019},\textbf{2020},\textbf{2021},\textbf{2022}},
      xtick=data,
      grid=major,
      grid style={dashed,gray!30},
      bar width=8,
      x tick label style={rotate=45,anchor=east},
      ylabel=Number of papers ,
      ylabel=Number of papers ,
      ]
    \addplot[ybar,fill=blue] coordinates { (\textbf{2019}, 6)(\textbf{2020}, 24) (\textbf{2021}, 49)    (\textbf{2022}, 45)};
\end{axis}
\end{tikzpicture}
\caption{SRGNN}
\label{fig:srgnn}
\end{subfigure}
\hfill
\begin{subfigure}{.25\linewidth}
\centering
\begin{tikzpicture}[scale=0.47]
\begin{axis}[
    symbolic x coords={\textbf{2019},\textbf{2020},\textbf{2021},\textbf{2022}},
      xtick=data,
      grid=major,
      grid style={dashed,gray!30},
      bar width=8,
      x tick label style={rotate=45,anchor=east},
      ylabel=Number of papers ,
      ylabel=Number of papers ,
      ]
    \addplot[ybar,fill=blue] coordinates { (\textbf{2019}, 2)(\textbf{2020}, 16) (\textbf{2021}, 41)    (\textbf{2022}, 53)};
\end{axis}
\end{tikzpicture}
\caption{NGCF}
\label{fig:ngcf}
\end{subfigure}
\hfill
\begin{subfigure}{.25\linewidth}
\centering
\begin{tikzpicture}[scale=0.47]
\begin{axis}[
    symbolic x coords={\textbf{2019},\textbf{2020},\textbf{2021},\textbf{2022}},
      xtick=data,
      grid=major,
      grid style={dashed,gray!30},
      bar width=8,
      x tick label style={rotate=45,anchor=east},
      ylabel=Number of papers ,
      ylabel=Number of papers ,
      ]
    \addplot[ybar,fill=blue] coordinates { (\textbf{2019}, 15)(\textbf{2020}, 28) (\textbf{2021}, 36)    (\textbf{2022}, 28)};
\end{axis}
\end{tikzpicture}
\caption{STAMP}
\label{fig:stamp}
\end{subfigure}
\hfill
\begin{subfigure}{.25\linewidth}
\centering
\begin{tikzpicture}[scale=0.47]
\begin{axis}[
    symbolic x coords={\textbf{2019},\textbf{2020},\textbf{2021},\textbf{2022}},
      xtick=data,
      grid=major,
      grid style={dashed,gray!30},
      bar width=8,
      x tick label style={rotate=45,anchor=east},
      ylabel=Number of papers ,
      ylabel=Number of papers ,
      ]
    \addplot[ybar,fill=blue] coordinates { (\textbf{2019}, 1)(\textbf{2020}, 6) (\textbf{2021}, 34)    (\textbf{2022}, 42)};
\end{axis}
\end{tikzpicture}
\caption{BERT4REC}
\label{fig:bert4rec}
\end{subfigure}
\hfill
\begin{subfigure}{.25\linewidth}
\centering
\begin{tikzpicture}[scale=0.47]
\begin{axis}[
    symbolic x coords={\textbf{2016},\textbf{2017},\textbf{2018},\textbf{2019},\textbf{2020},\textbf{2021},\textbf{2022}},
      xtick=data,
      grid=major,
      grid style={dashed,gray!30},
      bar width=8,
      x tick label style={rotate=45,anchor=east},
      ylabel=Number of papers ,
      ylabel=Number of papers ,
      ]
    \addplot[ybar,fill=blue] coordinates { (\textbf{2016}, 2)(\textbf{2017}, 5) (\textbf{2018}, 8)  (\textbf{2019}, 6)(\textbf{2020}, 12) (\textbf{2021}, 15)    (\textbf{2022}, 17)};
\end{axis}
\end{tikzpicture}
\caption{MF}
\label{fig:mf}
\end{subfigure}

\caption{The figures show the distribution over the years of the number of papers in which one of the top-15 most popular baselines was included.}
\label{fig:popular_dist}
\end{figure*}

\begin{table*}[t!]
     \caption{Performance comparison on RecBaselines2023. The best value is 	\textbf{in bold}.}
    \label{tab:benchmarkresults}
    \begin{tabular}{lcccccc}
    \toprule
     Model&R@10&R@20&N@10&N@20&M@10&M@20 \\ 
    \midrule
     
Random&0.045&0.08&0.029&0.029&0.008&0.008\\
BPRMF \cite{bpr}&0.2281&0.353&0.134&0.134&0.035&0.035\\
MostPop&0.312&0.339&0.138&0.138&0.035&0.035\\
MF2020 \cite{rendle2020neural}&0.348&0.446&0.227&0.227&0.067&0.067\\
EASER \cite{easer}&0.397&0.549&0.243&0.243&0.069&0.069\\
NeuMF \cite{neumf}&0.42&0.513&0.252&0.252&0.073&0.073\\
Slim \cite{slim}&0.446&0.576&0.264&0.264&0.078&0.078\\
VAECF \cite{multvae}&0.455&0.603&0.264&0.264&0.075&0.075\\
$RP3_{\beta}$ \cite{rp3beta}&\textbf{0.473}&\textbf{0.607}&\textbf{0.303}&\textbf{0.303}&\textbf{0.088}&\textbf{0.088}\\

       \hline
    \end{tabular}
    \end{table*} 

\autoref{fig:dataset_dist12} and \autoref{fig:dataset_dist3} represent three distributions of the dataset: the distribution of the number of papers over the year they were published, the distribution of the number of papers over the number of algorithms included in a paper, and the distribution of the number of papers for the top 10 most popular baselines included in our dataset. The earliest publication date of a paper is 2009, the number of papers remains relatively small and only exceeds 10 in 2017. Then the number of papers increases significantly from year to year. As can be seen in \autoref{fig:dataset_dist12}, a typical number of baselines included is between 3 and 8. Therefore, the algorithms used to recommend baselines for recommender systems have to work with a small number of available interactions. 

\autoref{fig:popular_dist} describes the distribution of the number of papers over the years for each of the top 15 popular baselines from the RecBaselines2023 dataset. The most popular models are BPR, GRU4Rec, LightGCN, NeuMF and others. These models were used as starting points for the collection of other papers. Therefore, they are represented in the dataset in large numbers.

\section{Collaborative Filtering for Baseline Selection}
Baseline selection can be solved by collaborative filtering (CF) algorithms. For example, the following definition was given in \cite{schafer2007collaborative}.

\begin{definition}
Collaborative filtering is the process of filtering or ranking items using the opinions of other people. 
\end{definition}

We can replace the word "items" with "baselines" and the word "people" with "researchers". This definition then provides a justification for the use of the technique. In addition, scientists' "opinions" are often motivated by several reasons. The first is the desire to compare the new algorithm with the best-known or best-performing approaches. The second is to include models based on the same idea. For example, the authors of \cite{shen2021powerful} compare their graph-based model with 4 baselines, 3 of which are also graph-based. These or other reasons explain the choice of models from a large number of options.

Therefore, researchers and practitioners may be interested in baseline recommendations based on a partial list of algorithms already in use. Hopefully, this can be done by applying approaches in the inductive scenario \cite{schnabel2022situating}. These approaches do not have ID-based user embeddings \cite{shen2021powerful, multivae, wu2022inmo, ananyeva2022towards}. They understand user interests based on the set of interactions. Therefore, we can easily adopt such techniques for suggesting baselines based on a partial list of already included methods.

\section{Experiments}
We have experimented with collaborative filtering approaches on the Top-N recommendation task on the RecBaselines2023 dataset. Our experiments aim to answer the following question: "What is the performance of different state-of-the-art collaborative filtering approaches on the RecBaselines2023 dataset?"

\textbf{Models}. We included popular approaches of different types: simple random and MostPop; matrix factorisation based BPRMF \cite{bpr}, MF2020; item based EASE \cite{easer}, SLIM \cite{slim}; graph based $RB3$ \cite{rp3beta}, VAE based MultiVAE \cite{multivae}. According to \cite{topnsota, ferrari2019we}, such models are very strong CF-based baselines.


\textbf{Metrics}. Standard quality ranking metrics are chosen, namely Recall@K, NDCG@K and MAP@K.

\textbf{Experiment settings}. To provide reproducible experiments we use Elliot \cite{elliot} similar to \cite{topnsota}. This framework allows experiments to be fully described in a configuration file. This file is available online \footnote{\url{https://github.com/fotol1/recbaselines2023}} and hyperparameter ranges can be found there. The total number of hyperparameters set for each model is 20.

\textbf{Evaluation protocol}. All interactions are divided into train/valid/test splits. The valid split is used for early stopping and hyperparameter selection. The final quality is estimated on the test split. All papers published before 2021 are used for training. In addition, 80 $\%$ of the interactions for papers published in 2021 and 2023 are used for training, and the remaining 20 $\%$ are used for validation and test, respectively.

\textbf{Results}. To investigate our question, we report quality metrics for different approaches in \autoref{tab:benchmarkresults}. As we can see, the best model $RP3_{\beta}$ is two times better than MostPop's recommendations. This shows that there are not many universal baselines in recommendations, and that researchers choose baselines carefully. Surprisingly, the best model, $RP3_{\beta}$, has a Recall@20 of 0.6. 
This means that we can find more than half of the hidden baselines in lists of length 20.

\begin{figure*}
     \centering
     \includegraphics[width=0.8\textwidth]{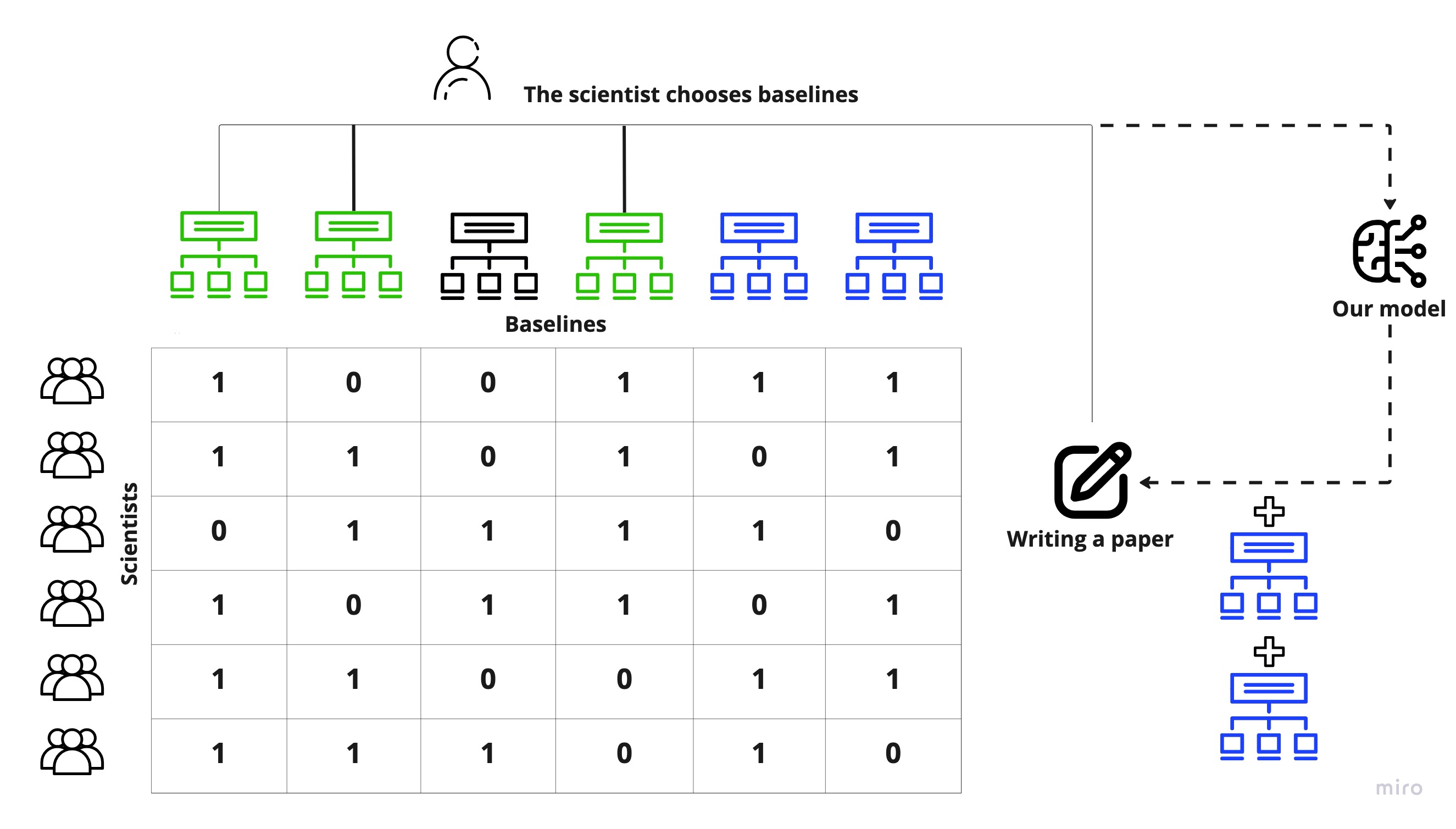}
     \caption{The scientist chooses baselines for his experiments. Three baselines have already been chosen. He or she can now pass the three approaches to one of the collaborative filtering algorithms under consideration. As a result, two more baselines are suggested as additional approaches to include in the paper. Based on historical data, these recommended baselines may also be chosen by other researchers.}
     \label{fig:main_schema}

\end{figure*}

\section{Selecting baselines for partial lists}

This section describes one possible way of using RecBaselines2023. The \autoref{fig:main_schema} represents the main idea. A scientist has invented a new recommendation algorithm and wants to compare it with other work. For example, a new approach was inspired by two methods \textbf{a} and \textbf{b}. So they are automatically included in the experiments of the new paper. In addition, the researchers know that a current state-of-the-art algorithm is a model \textbf{c}. So it should also be considered. Given the set of three baselines $\{\textbf{a}, \textbf{b}, \textbf{c}\}$, he or she can run the set in one of the adapted collaborative filtering approaches. This will return a list of recommended baselines. The choice of these is consistent with the historical data represented in RecBaselines2023. 

\begin{table*}[t!]
\caption{Examples of recommendations based on partial lists of items.}
\label{tab:examples}
\resizebox{\textwidth}{!}{%
    \begin{tabular}{lccc}
    \toprule
    \centering
     Input items & SLIM  & EASE & $RP3_{\beta}$ \\ 
    \midrule
    
    GRU4REC&CASER, SASREC&SASREC, CASER&MIND, DIEN\\ 
    GRU4REC, SASREC, BERT4REC&CASER, TISASREC&BPR, CASER&TISASREC, JODIE\\ 
    GRU4REC, SASREC, BERT4REC, TISASREC&CASER, S3REC&BPR, CASER&LESSR, CHORUS\\ 
    \midrule
    PINSAGE&GCMC, CMN&NGCF, GCMC&CMN, GCMC\\ 
    VAECF&TRANSCF, LRML&LIGHTGCN, TRANSCF&SGL, TRANSCF\\ 
    RIPPLENET&PER, CKE&CKE, PER&PER, LIBFM\\ 
    LIGHTGCN, SGL, VAECF&TRANSCF, LRML&NGCF, BPR&TRANSCF, SBPR\\ 
    
    \hline
    \end{tabular}
}
\end{table*}
In \autoref{tab:examples} we demonstrate recommendations for some sub-lists of baselines. We use SLIM, EASE and $RP3_{\beta}$ as recommender models because they are item-based models that can make predictions based on any input list of items. The first three examples emulate iterative updates to the next-item recommender set of baselines. The next examples demonstrate recommendations based on a single baseline using different frameworks. As we can see, SLIM and $RP3_{\beta}$ are flexible in changing recommendations as new next-item models are added. When we provide only one element of a particular framework, our models recommend baselines using similar frameworks. For example, RippleNet \cite{ripplenet} is a knowledge-based model. If someone includes RippleNet in their experiments, our models will suggest including other knowledge-based approaches such as PER \cite{per}, CKE \cite{cke}.

\section{Limitations and future work}
Our work has some limitations. In this section we will discuss them and show possible ways to overcome them.

Firstly, the published version of the dataset will become obsolete. We will publish regular updates. In addition, if authors of newly proposed methods want to add their work, we can do this quickly in the repository via a pull request.

Secondly, The dataset may contain misspellings or other errors, with the presence or absence of some baselines in the included works. We have tried to do our best, and have double-checked the interactions several times. If you find any errors, please contact us via issues on Github.

Finally, There are some challenges in publishing baseline recommendations. For example, some of the baselines presented have been used in previous work. However, the latest state-of-the-art approaches replace them. The models considered are not sensitive to this fact. We argue that this problem exists for other datasets as well. It has been shown in \cite{toppopular} that recommending the most recent films can improve quality even for the simple MostPop method. Nevertheless, the practical application can be modified and the most recent baselines with high relevance scores can be treated as more suitable. We leave this as future work.

\section{Conclusion}
This paper investigates the problem of recommending baselines for experiments. We have collected an open source dataset RecBaselines2023, which describes baseline models used for comparative experiments in papers on different types of recommender systems. It consists of 903 papers and 363 baseline models, with 5467 interactions between them. The dataset includes interactions between papers and baseline models, and additional data about each paper, such as web link to a paper, paper title, and year of publication. RecBaselines2023 can be used by researchers to properly compile the baseline list for their experiments. The dataset will be updated as new papers are published.  We have used collaborative filtering techniques to identify the best algorithms based on incomplete lists of previously included baselines. Our experiments with hidden predictions of recommender baselines show that state-of-the-art collaborative filtering techniques can successfully perform this task. We hope that our dataset can open up new lines of research.

\bibliographystyle{ACM-Reference-Format}
\bibliography{main}


\end{document}